\documentclass{aa}  
\usepackage{graphicx}
\usepackage{txfonts}
\usepackage{hyperref}
\usepackage{soul}
\usepackage{multirow} 
\usepackage{float}
\usepackage{orcidlink}
\usepackage{xcolor}
\usepackage{natbib}

\DeclareRobustCommand{\VAN}[3]{#2}
\let\VANthebibliography\thebibliography
\def\thebibliography{\DeclareRobustCommand{\VAN}[3]{##3}\VANthebibliography}
\usepackage{longtable}
\newcommand{\angstrom}{\textup{\AA}}

\usepackage[switch]{lineno}
\usepackage[normalem]{ulem}

\begin{document}

%\linenumbers

\title{Multi‑epoch spectroscopic variability of the B supergiant HD\,75149}

   \author{J. Chamoun-Contreras \orcidlink{0009-0000-2113-3521}
          \inst{1}
          \and
          C. Arcos
          \inst{2} \orcidlink{0000-0002-4825-4910}
          \and
         N. Machuca          
         \inst{2} \orcidlink{0000-0001-5638-1437}
         \and
         C. E. Perez-Ramirez          
         \inst{3} \orcidlink{0009-0001-6355-0371}
         \and
         L. S. Cidale 
         \inst{4,5} \orcidlink{0000-0003-2160-7146}
         \and
          M. Curé 
          \inst{2} \orcidlink{0000-0002-2191-8692}
          \and
         I. Araya
          \inst{6} \orcidlink{0000-0002-8717-7858}
          \and
         D. Turis-Gallo
          \inst{2} \orcidlink{0009-0000-4835-3479}
          \and
          M. Hadjara
          \inst{7,8} \orcidlink{0009-0002-6955-8443}
          }
   
       \institute{Centro de Astronomía (CITEVA), Universidad de Antofagasta Av. Angamos 601, Antofagasta, Chile
         \and
             Instituto de F\'{\i}sica y Astronom\'{\i}a, Facultad de Ciencias, Universidad de Valpara\'{\i}so,
             Av. Gran Breta\~na 1111, Valpara\'{\i}so, Chile
         \and
         Departamento de Computación e Industrias, Universidad Católica del Maule, Avenida San Miguel 3605, Talca, Chile.
         \and
         Departamento de Espectroscop\'{\i}a, Facultad de Ciencias Astron\'omicas y Geof\'{\i}sicas, Universidad Nacional de La Plata, Paseo del Bosque S/N, BF1900FWA La Plata, Buenos Aires, Argentina
         \and
          Instituto de Astrof\'{\i}sica de La Plata, CCT La Plata, CONICET-UNLP, Paseo del Bosque S/N, BF1900FWA La Plata, Buenos Aires, Argentina
          \and
          Centro Multidisciplinario de F\'isica, Vicerrector\'ia de Investigaci\'on, Universidad Mayor, 8580745 Santiago, Chile
          \and
          Nanjing Institute of Astronomical Optics \& Technology, National Astronomical Observatories, Chinese Academy of Sciences, Nanjing 210042, China
          \and
            Chinese High Angular Resolution Southern Astronomical Laboratory (CHARSAL)/Astro-Photonics Laboratory/Space and Planetary Exploration Laboratory (SPEL), Dept. of Electrical Engineering, FCFM, Universidad de Chile, Av. Tupper 2007, Santiago, Chile
            }

\titlerunning{Variability of the B supergiant HD\,75149}
\authorrunning{J. Chamoun-Contreras et al.}
\date{Received: 17/11/2025; Accepted: 26/04/2026 }

  \abstract 
   {Massive stars continuously enrich the surrounding interstellar medium by supplying it with stellar material driven by their powerful winds. B supergiant stars (BSGs) in particular are a type of massive star characterized by strong winds and notable photometric and spectroscopic variability.}
   {We aim to conduct a pilot study of the optical spectroscopic variability of the BSG HD\,75149 between 2004 and 2025. Its extended temporal baseline and pronounced variability amplitude make it particularly well suited for investigating the physical origin of the observed short-term variability within a consistent hydrodynamical and radiative-transfer framework.}
   {We analyzed 25 nightly averaged optical spectra obtained with different instruments and telescopes, some of them with observations over several consecutive days. We measured the radial velocities (RVs) and equivalent widths (EWs) of 17 spectral lines (H, \ion{He}{I}, \ion{Si}{III}, \ion{N}{II}, \ion{Mg}{II}, \ion{C}{II}). We modeled the H$\alpha$ emission, absorption, and P-Cygni profiles using the ISOSCELES grid and the $\delta$-slow hydrodynamic regime.}
   {H$\alpha$ shows variability in intervals of a few days, including P-Cygni changes, while metal lines show small RV amplitudes, consistent with pulsating oscillations. The largest variation in the mass-loss rate corresponds to an increase of a factor of 1.8 within four days. In contrast, the terminal velocity remains barely affected during the same time interval. }
   {The pronounced variation observed in hydrogen lines, in contrast with the variability of other lines, suggests that it is due to mass-loss rate episodes driven by a slow wind occurring on a timescale comparable to photometric variations. We found no evidence of a close binary companion in the sample used, but we cannot completely exclude the possibility of a wide or low-inclination companion.}


   \maketitle

\section{Introduction}
Massive stars lose material through powerful stellar winds that shape their evolution and enrich the interstellar medium both chemically and dynamically. On a galactic scale, this process has a significant influence on the environment, particularly when these stars end their lives as supernovae. Among them, blue supergiant stars (BSGs) display particularly complex spectroscopic and photometric behavior, often showing variability in line broadening, flux intensity, radial velocities (RVs), and equivalent widths (EWs; \citealt{simon2024}). Unlike cooler evolved stars, such as red supergiants (RSGs) or asymptotic giant branch (AGB) stars, BSGs do not develop an extended molecular sphere (“molsphere”) due to their high effective temperatures \citep{Hadjara2019}. These variations have given rise to multiple interpretations of their physical origin.

Spectroscopic variability in BSGs is influenced by temporal variations in the stellar wind, which modulate the mass-loss rate and shape the observed line profiles \citep{kraus2015,Cidale2023}. The modified Castor--Abbott--Klein (m-CAK) theory \citet{Castor1974,cak1975,Abbott1982,Friend1986,PPK1986} describes the line-driven winds and governs the wind acceleration, determining the mass-loss rate ($\dot{M}$) and terminal velocity ($v_\infty$) of the wind (see the review by \citealt{Cure2023}).

\citet{Cure2011} identified the $\delta$-slow solution, a new hydrodynamic solution that provides a possible explanation for the slow winds observed in A-type supergiants and B hypergiants \citep{Venero2024}.
Building on this framework, the ISOSCELES model grid of \citet{araya2025} incorporates both fast and $\delta$-slow wind solutions in spectral analyses, replacing the traditional $\beta$-law velocity prescription. Initial applications to BSGs show that the $\delta$-slow solution provides an improved reproduction of observed spectral features compared to standard wind models \citep{Venero2024,ortiz2025}.

These advances in wind modeling naturally raise the question of what physical mechanisms are responsible for the variability observed in massive stars. Although magnetism has been ruled out as the main driver of variability in massive stars \citep{scholler2017}, binarity has emerged as a key factor influencing both stellar evolution and observed variability. Recent large-scale surveys have provided new insight into the multiplicity of BSGs. \citet{Britavskiy2025}, within the Binarity at Low Metallicity (BLOeM) campaign, analyzed 262 early-type BSGs (B0–B3) in the Small Magellanic Cloud and found a spectroscopic binary fraction of $34\,\pm\,3\,\% $. For later-type BSGs (B5–B9), \citet{Patrick2025} reported an apparent binary fraction of $25\,\pm\,6\,\% $, which decreased to $ <18\,\%$ after accounting for observational biases. This reduction highlights how intrinsic stellar variability, particularly pulsations typical of $\alpha$ Cyg stars, can mimic binary-induced RV variations.

Periodic variations in RV caused by orbital motion are generally detected through the shift of metallic lines (e.g., silicon and carbon) relative to their rest wavelengths \citep[see Gaussian methods in][]{Moravveji2012}. However, intrinsic pulsations also play a major role in the variability of BSGs.

The dominant pulsation modes in these stars are pressure (p), gravity (g), and strange (s) modes. P-modes are acoustic oscillations for which pressure acts as the main restoring force, with periods of hours to about one day. G-modes are internal gravity waves restored by buoyancy \citep[e.g.,][]{Aerts2021}, with periods from one to 10 days. S-modes arise from instabilities in highly nonadiabatic, radiation-pressure-dominated outer layers, often associated with opacity effects \citep[e.g.,][]{Georgy2021,Agrawal2022}. Periods typically range from ten to 100 days. 
Despite these theoretical period values, \cite{lefever2007} found a period of 2.7 h for the B8 Ib BSG HD\,46769, and \cite{kraus2012} reported an even shorter period of 1.59 h for the B9 Iab HD\,202850, imposing a new discussion about the changes in the internal structures of late BSGs.

We focused on HD\,75149 due to its complex spectroscopic variability within an extended temporal baseline (see also \citealt{Haucke2018}). 
This star, also known as OP Vel, is a BSG of spectral type B3\,Ia located in the Vela nebula at a distance of 1771.48 pc, with a visual magnitude of 5.47 mag \citep{Gaia2018}. \citet{Haucke2018} modeled the Balmer, helium, and silicon line profiles with the non-local thermodynamic equilibrium (non-LTE) atmosphere code FASTWIND \citep{Puls2005}, obtaining an effective gravity of $\rm \log g\,=\,2.12\,\pm\,0.05$, an effective temperature of $\rm T_{eff}\,=\,16\,000\,K$, a stellar radius of $\rm R_{\ast} =\,61\,\pm\,13 \,R_{\odot}$, and a projected rotational velocity of $\rm v\sin i\,=\,40\,km\,s^{-1}$. In comparison, \citet{Kourniotis2025} derived $\rm \log g\,=\,2.5$, $\rm T_{eff}\,=\,15\,849\,K$, $\rm R_{\ast}\,=\,50\,R_{\odot}$, and $\rm v\sin i\,=\,30\,km\,s^{-1}$.

Photometric studies have long revealed variability in HD\,75149. Using Hipparcos data, \citet{Koen&Eyer2002} detected a 1.086 d period, classified as “unsolved.” Later, \citet{lefevre2009} reclassified the star as a slow pulsating B-type (SPB) star, while \citet{lefever2007} identified multiple periods (1.215–2.214 d) consistent with $\alpha$ Cyg variability. From TESS data (Sectors 8+9, 35+36, and 62), \citet{Kourniotis2025} obtained independent frequencies ranging between 3.228 and 8.570 days using iterative pre-whitening. A summary of the reported periods is provided in Table \ref{tab:data}.

\begin{table}[h!]
        \centering
    \caption{Periods and variability classification from the literature.}
        \label{tab:data}
        \begin{tabular}{l c c}
                \hline
                \hline
        Method  &       Period (days)   &       Var. Type \\
        \hline
    LC (Hipparcos) &    1.086$^{a}$     &       Unsolved \\
    LC (Hipparcos) &    1.086$^{b}$     &       SPB\\
    LC $\&$ Spectroscopy &      1.215 -- 2.214$^{c}$    &       $\alpha$ Cyg \\
    LC (TESS) S35-S36   &       5.356 -- 3.228 -- 4.579$^{d}$   &       --\\
    LC (TESS) S8-S9     &       6.151 -- 4.889$^{d}$    &       --\\
    LC (TESS) S62 &     6.120 -- 8.570$^{d}$    &       --\\
                \hline
                % \hline
        \end{tabular}
\tablefoot{LC stands for light curve. References: $^{a}$~\cite{Koen&Eyer2002}; $^{b}$~\cite{lefevre2009}; $^{c}$~\cite{lefever2007}; $^{d}$~\cite{Kourniotis2025}.}
\end{table} 

Through consecutive spectroscopic observations of HD\,75149, we aim to investigate the origin of this variability by exploring the possible roles of binary and stellar wind properties. Other early BSGs have also presented strong night-to-night variability, for example $\rho$ leo, $\epsilon$ Ori, HD\,14134, HD\,37128, 55 Cyg, and HD\,198478  (\citet{morel2004,kraus2012,kraus2015,tomic2015,Cidale2023,checha2026}), but mass-loss-rate determination through time only exists for a few, for example HD\,75149, HD\,53138, and HD\,111973 \citet{Haucke2018}. Moreover, this is the first work to apply the slow-wind solution to consecutive observations to study wind variability. This work has two main goals: (i) to investigate the short-term spectroscopic variability of HD 75149 using observations collected on consecutive days, and (ii) to evaluate the diagnostic power of high-cadence spectroscopy combined with hydrodynamic wind models in the study of BSGs. \\

In Sect.~\ref{sec:2} we describe the acquisition and reduction steps of spectroscopic data and the averaging criteria for the spectra observed during the same night. In Sect.~\ref{sec:3} we explain the method implemented to obtain the RV and EW measurements and the criteria used to analyze their variability. In Sect.~\ref{sec:4} we describe our study of the stellar wind variability from H$\alpha$ using the ISOSCELES database to derive $\rm\dot{M}$ and $\rm v_{\infty}$. Section~\ref{sec:5} discusses the results in the context of stellar wind variability and the probability of binarity, integrating evidence from photometric observations of previous work in the literature. Finally, Sect.~\ref{sec:6} summarizes the main conclusions, and Appendix~\ref{appendixA} compiles the complete RV and EW tables and supporting material.

\section{Data acquisition}\label{sec:2}
For this study, we used 25 averaged optical spectra from two different instruments and observatory facilities. These are listed below. 

\begin{enumerate}
    \item[\Roman{enumi}i.] Nineteen observations were performed between January 2006 and February 2017 \citep{Haucke2018} using the REOSC échelle spectrograph in cross dispersion (DC) mounted at the 2.15 m Jorge Sahade (JS) telescope in the Complejo Astronómico El Leoncito (CASLEO)\footnote{\url{https://casleo.conicet.gov.ar/}}, San Juan, Argentina. The instrument covers a spectral range of 4300–6800 $\AA$ and has a resolving power of around 13,000. Data reduction followed standard procedures using the NOAO/IRAF package\footnote{The Image Reduction and Analysis Facility (IRAF), developed and supported by the National Optical Astronomy Observatories (NOAO) in Tucson, Arizona, operated by the Association of Universities for Research in Astronomy (AURA).}, including wavelength calibration, which resulted in an average signal-to-noise ratio (S/N) of approximately 300. For details about instrumentation setup and reduction steps, see \citet{Haucke2018}. For these observations, we adopted a single-epoch RV uncertainty of 3–5 $\rm km\,s^{-1}$ and a 10\% uncertainty in EW for the REOSC-DC data \citep{naze2016}.  
    \item[\Roman{enumi}ii.] Two observations were performed in January 2025 with the Las Cumbres Observatory Network of Robotic Échelle Spectrographs (NRES/LCO) with a resolving power of $\sim$53\,000 mounted on 1m telescopes, under the program allocated by the Chilean Telescope Allocation Committee (CNTAC), N°: [CN2024B-6]. For data reduction, we utilized the open-source BANZAI pipeline, for which details about the standard steps are available online \footnote{\url{https://github.com/LCOGT/banzai-nres}}. The estimated S/N of the final spectra is approximately 100. At this spectral resolution and S/N ratio, we adopted a single-epoch RV precision of $\approx$ 10 $\rm m\,s^{-1}$ ($\approx$ 0.01 $\rm km\,s^{-1}$) for NRES estimated by \cite{Brandt2020} using the \textit{xwavecal} wavelength calibration algorithm.
    \item[\Roman{enumi}iii.] Four spectra were observed between December 2004 and February 2006 using the fiber-fed extended-range optical spectrograph (FEROS) mounted on the MPG/ESO 2.2 m telescope at La Silla (Chile). FEROS is a bench-mounted, fiber-fed échelle spectrograph with cross-dispersion by prisms. The data were reduced with the ESO-MIDAS FEROS DRS and provided to us as merged and barycentrically corrected 1D spectra with a spectral resolution of R=48,000, single-order S/N values of $\approx$ 200-230, and a coverage of 3500--9200 $\AA$. For this FEROS data, we adopted a single-epoch RV uncertainty of 14.5 $\rm m\,s^{-1}$ ($\approx$ 0.015 $\rm km\,s^{-1}$ ) reported by \cite{Trifonov2018}.

\end{enumerate}

To place all RV measurements on a homogeneous reference frame and to quantify instrumental systematic effects, we treated the wavelength solutions in a consistent way for the three datasets. The FEROS spectra, reduced with the FEROS-DRS pipeline, provide a barycentric correction on output \citep{Soto2015}; however, because shortcomings reported in the literature \citep[including the use of an inappropriate time stamp;][]{Muller2013,Soto2015}, we recomputed the barycentric correction for all FEROS, REOSC-DC, and NRES spectra at mid-exposure and applied it to the wavelength scale using the target ICRS coordinates and the observatory location \citep{Stumpff1980,WrightEastman2014}. After this step, the dominant instrumental systematics are residual-wavelength zero-point shifts and (for fiber-fed data) illumination and flexure effects \citep{Baranne1996,Munari1992}. We therefore estimated a per-spectrum zero-point term by measuring the apparent velocity of narrow non-stellar features within our wavelength range (telluric O$_2$ lines around $\sim$6277\,\AA\ when present \citep{Figueira2010} and/or strong diffuse interstellar bands, which are of interstellar origin and thus largely insensitive to the stellar line-profile variability \citep[e.g.,][]{Herbig1995,Hobbs2008}), and we subtracted this offset from the stellar RVs. The remaining inter-instrument offsets are consistent with zero within the REOSC-DC calibration floor inferred from the same zero-point tracers; no additional global RV offset correction is needed. In practice, the RV precision achieved in this work is not limited by the intrinsic stability expected for fiber-fed facilities (i.e., FEROS) or targeted by modern robotic fiber-fed échelle systems (i.e., NRES), but by the width and time-variable asymmetries of the photospheric and wind-formed lines in a B supergiant \citep[e.g.,][]{Siverd2018,Sterken1978,Aerts2009,SimonDiaz2017}; we therefore report RV uncertainties from our line-center estimator (Sect.~3), and we summarize the typical single-epoch precision per instrument by the median (and 16th--84th percentile) of the internal RV errors; these were measured across the photospheric metal lines for each dataset.

Before averaging the optical spectra, we examined short-term variability by analyzing one- to three-hour intervals and found no significant changes. Consequently, we combined the spectra from each night, yielding a final sample of 25. For each spectrum, we measured the EWs and RVs of 17 spectral lines, as listed in Table~\ref{tab:LabW}.

\begin{table}[h!]
        \centering
    \caption{Lines at rest air wavelength used in this work.}
        \label{tab:LabW}
        \begin{tabular}{l c} 
                \hline
                \hline
        Line    & Wavelength \\
            &   [$\AA$] \\
               \hline
H$\alpha$ & 6562.71 \\
H$\beta$ & 4861.28 \\
H$\gamma$ & 4340.47 \\
\ion{Si}{III} & 4552.62 \\
& 4567.82 \\
& 4574.76\\
HeI & 4387.93\\
& 4471.48\\
& 5015.68\\
& 5875.62\\
\ion{N}{II} & 4601.48\\
& 4607.16\\
& 4630.54\\
& 5666.63\\
\ion{Mg}{II} & 4481.13\\
\ion{C}{II} & 6578.05\\
& 6582.88\\
                \hline
        \end{tabular}
\tablefoot{The values are taken from NIST Atomic Spectra Database Lines \citep{NIST}.}
\end{table} 

\section{Radial-velocity and EW variability}\label{sec:3}
The time sampling of the spectra obtained in 2013, 2014, 2015, and 2025 allowed the analysis of variations on timescales of a few days and the search for trends over consecutive months. Table~\ref{tab:tracts} summarizes the observational dates and corresponding Julian day intervals used in this work. Unfortunately, the time sampling of the spectra obtained in 2004, 2006, and 2017 is too sparse for such an analysis.

\begin{table}[h!]
        \centering
    \caption{Time intervals studied in Julian days.}
        \label{tab:tracts}
        \begin{tabular}{l l} 
                \hline
                \hline
        Time interval & Date range\\
    (JD 2400000+)       &  \\
        \hline
    56329.5 - 56331.7 & Feb 2013\\
    56740.6 - 56741.6 & Mar 2014\\
    56758.5 - 56761.5 & Apr 2014\\
    57067.5 - 57069.5 & Feb 2015\\
    57095.5 - 57098.6 & Mar 2015\\
    60698.4 - 60700.7 & Jan 2025\\
                \hline
        \end{tabular}
    \tablefoot{This table does not consider single observations since the aim is to analyze the variation in the order of days for the magnitudes studied.}

\end{table}

\begin{figure*}
    \centering
    \includegraphics[width=0.44\linewidth]{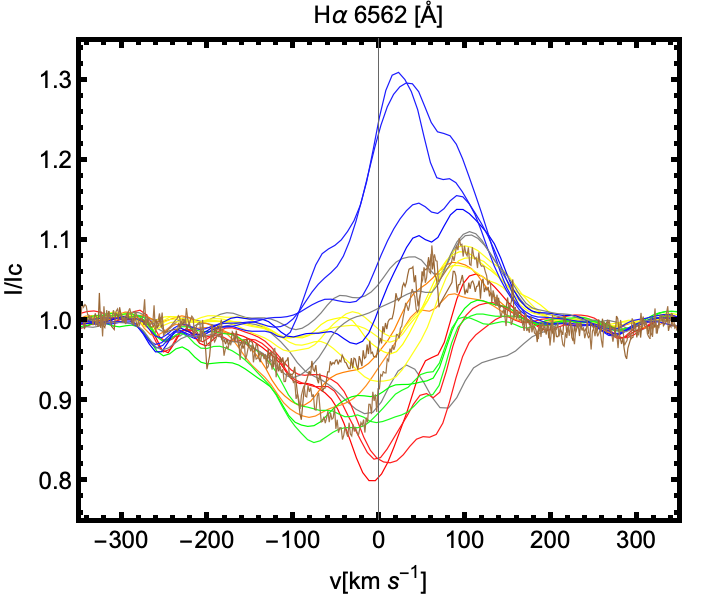}
    \includegraphics[width=0.44\linewidth]{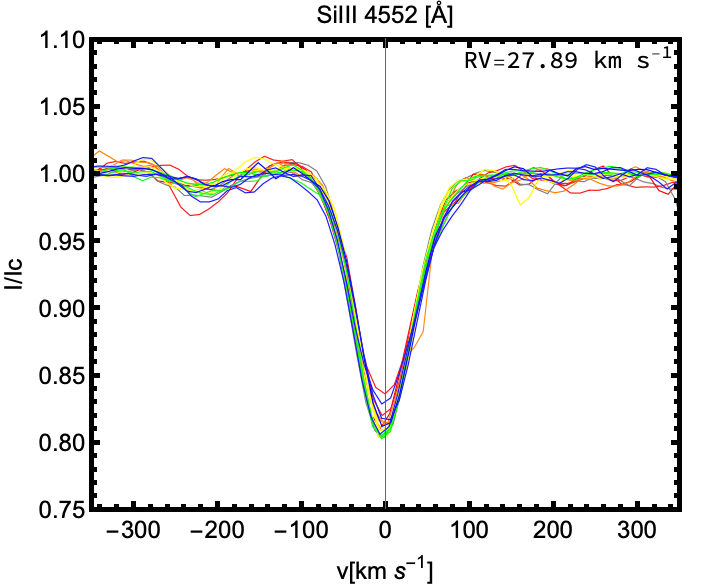}
    \caption{\textit{Left:} H$\alpha$ emission, absorption, and P-Cygni profiles over the years 2004–2025. The star exhibits short-term variability on a daily basis.  \textit{Right:} \ion{Si}{III} 4552 $\angstrom$ absorption line for the same period but without the spectrum of 2025. The spectral lines are shifted 27.89 $\rm km\,s^{-1}$ from the central position. No significant intensity variation is observed for this atomic line. For both images, a color code is assigned to spectra belonging to the same data range described in Table~\ref{tab:tracts}; for the three spectra outside these ranges, gray is used. For instance, red is for Feb 2013, orange for Mar 2014, yellow for Apr 2014, green for Feb 2015, blue for Mar 2015, and brown for Jan 2025.}
    \label{fig:all_spectra}
\end{figure*}

To highlight the variability observed in the H$\alpha$ P-Cygni profile, we plot the 25 spectra (from 2004 to 2025) in Fig.~\ref{fig:all_spectra}. Although the entire observation range is displayed, changes in the line profile shape occur over consecutive days; indeed, in some cases, the line shifts from absorption to complete emission within a single month (see Fig.~\ref{fig:rvha}). For the rest of the spectral lines, \ion{Si}{III}, \ion{He}{I}, \ion{N}{II}, \ion{Mg}{II}, and \ion{C}{II}, a smaller amplitude of spectral-line-profile variations is observed. In the right panel of Fig.~\ref{fig:all_spectra}, \ion{Si}{III}  4553 is displayed as an example. 

To analyze the origin of the line variability, we calculated the moments $  M_0  $ (equivalent width), $  M_1  $ (centroid), and $  M_3  $ (skewness) of each spectral line following \citealt[p. 240--41]{Asteroseismology2010}.
The uncertainty of each moment was estimated by repeating its calculation ten times. In each iteration, we progressively modified the wavelength range included in the sum by starting with the outermost boundaries of the confidence region (where the data unambiguously belong to the emission or absorption component) and gradually shrinking it inward toward the uncertainty region (around the continuum level, where it is unclear whether the points belong to the line). We always stopped before including points from the negative region, where the data were considered to belong neither to the line nor to a safe continuum. These three regions were defined using an approach inspired by type-1 fuzzy logic \citep{saatchi2024fuzzy}. The final value of each moment was taken as the mean of the ten iterations, and the associated uncertainty was estimated as the standard deviation of the resulting values.

Table~\ref{tab:ha_var} presents the RV and EW results for the H$\alpha$ emission line and its absorption component separately. Similarly, the complete set of RV and EW values, along with variations per time interval for the other spectral lines, is included in Tables~\ref{tab:RV_var} and~\ref{tab:EW_var}.

To assess the significance of the variability in RV, we distinguish between the intrinsic error in the line-center measurement and the instrumental floor for each dataset. In the REOSC-DC spectra, our manuscript adopts a typical uncertainty per epoch of 3–5 km/s, which is consistent with the literature for this instrument \citep{naze2016}; in what follows, we use 4 km/s as a representative value. Thus, when comparing two epochs, we define $\sigma_{\Delta}=\sqrt{\sigma_1^2+\sigma_2^2}$ and consider a RV variation to be probable when $|\Delta {\rm RV}| \geq 2\sigma_{\Delta}$, while reserving the criterion $|\Delta {\rm RV}| \geq 3\sigma_{\Delta}$ for robust cases. For two REOSC-DC measurements of comparable quality, this corresponds to thresholds of $\sim 11$ km/s and $\sim 17$ km/s, respectively. This choice is supported by our own measurements; across the photospheric metal lines, the internal RV uncertainties returned by the line-centering procedure have a median of 0.31 km\,s$^{-1}$, with a 16th--84th percentile range of 0.15--0.64 km\,s$^{-1}$, which is well below the adopted instrumental floor. Therefore, for REOSC-DC, the relevant error budget is dominated by the instrumental precision rather than on the noise of the centroid \citep{Bouchy2001}.

The EW uncertainties cited can be directly compared with the errors obtained using our measurement procedure \citep{VollmannEversberg2006}. For REOSC-DC spectra (\(S/N\approx300\)), the iterative integration scheme yields, for photospheric metal lines, a median absolute uncertainty of \(0.0020\)\,\AA, with an 84th percentile of \(0.0042\)\,\AA \, and a maximum of \(0.0081\)\,\AA. In relative terms, this corresponds to a median uncertainty of \(0.9\%\), an 84th percentile of \(1.8\%\), and a maximum of \(4.1\%\); these values are clearly lower than the typical errors of \(5\)–\(10\%\) in EW reported for REOSC-DC spectra. In absolute terms, a \(10\%\) error would be equivalent to approximately \(0.02\) - \(0.04\) \,\AA \, for most photospheric metal lines in our sample and could reach \(0.09\) - \(0.13\) \,\AA \, in the most intense lines, i.e., values clearly greater than the uncertainties derived from our iterative procedure. The Balmer lines exhibit somewhat larger absolute uncertainties, typically on the order of \(0.003\)--\(0.006\)\,\AA, with a maximum of \(0.023\)\,\AA \, for H\(\gamma\); however, their relative errors remain small, except for H\(\alpha\), whose EW can approach zero or change sign, making the percentage uncertainties unstable. Overall, these uncertainties are consistent with the spectra's quality and can be considered conservative.
The errors from REOSC-DC were taken as representative for comparison purposes since they exhibit greater uncertainty, and these spectra make up the majority of the sample analyzed.

\begin{table}[!ht]
    \centering
    \caption{Line-by-line statistics of RVs and EWs.}
    \small
    {\renewcommand{\arraystretch}{1.25}
    \begin{tabular}{lrrrrrr}
        \hline\hline
        Line\rule{0pt}{2.6ex} & $\overline{\mathrm{RV}}$ & $\overline{\mathrm{\Delta RV}}$ & $\sigma_{\mathrm{RV}}$ & $\overline{\mathrm{EW}}$ & $\overline{\mathrm{\Delta EW}}$ & $\sigma_{\mathrm{EW}}$ \\
        ~    & [{\rm km\,s$^{-1}$}] & [{\rm km\,s$^{-1}$}] & [{\rm km\,s$^{-1}$}] & [\AA] & [\AA] & [\AA] \\
        \hline
        H$\gamma$  & 22.39 & 0.397 & 10.16 & 1.48 & 0.006 & 0.32 \\
        \ion{He}{i} 4387 & 25.21 & 0.188 & 3.90 & 0.46 & 0.003 & 0.02 \\
        \ion{He}{i} 4471 & 22.34 & 0.121 & 3.45 & 0.78 & 0.003 & 0.05 \\
        \ion{Mg}{ii} 4481 & 21.83 & 0.211 & 3.87 & 0.33 & 0.002 & 0.03 \\
        \ion{Si}{iii} 4552 & 30.61 & 0.357 & 2.20 & 0.24 & 0.002 & 0.01 \\
        \ion{Si}{iii} 4567 & 31.35 & 0.366 & 3.60 & 0.19 & 0.002 & 0.02 \\
        \ion{Si}{iii} 4574 & 29.61 & 0.572 & 3.23 & 0.12 & 0.002 & 0.01 \\
        \ion{N}{ii} 4601 & 32.81 & 0.521 & 3.10 & 0.16 & 0.002 & 0.03 \\
        \ion{N}{ii} 4607 & 29.01 & 0.601 & 3.71 & 0.14 & 0.003 & 0.01 \\
        \ion{N}{ii} 4630 & 29.28 & 0.295 & 2.77 & 0.27 & 0.002 & 0.03 \\
        H$\beta$ & 17.56 & 0.254 & 6.34 & 1.27 & 0.006 & 0.20 \\
        \ion{He}{i} 5015 & 27.69 & 0.360 & 11.04 & 0.45 & 0.002 & 0.10 \\
        \ion{N}{ii} 5666 & 25.80 & 0.413 & 6.11 & 0.24 & 0.002 & 0.04 \\
        \ion{He}{i} 5875 & 27.01 & 0.092 & 4.46 & 0.89 & 0.002 & 0.08 \\
        \ion{C}{ii} 6578 & 30.32 & 0.494 & 5.57 & 0.34 & 0.003 & 0.04 \\
        \ion{C}{ii} 6582 & 24.72 & 0.272 & 2.63 & 0.23 & 0.004 & 0.03 \\
        \hline
    \end{tabular}}
    \label{tab:Mean}
\tablefoot{Summary by spectral line of RV and EW. For each line, $\overline{\mathrm{RV}}$ (average radial velocities), $\overline{\mathrm{\Delta RV}}$ (average uncertainties of individual measurements), and $\sigma_{\mathrm{RV}}$ (dispersion between observations) are reported. Similarly, $\overline{\mathrm{EW}}$, $\overline{\mathrm{\Delta EW}}$, and $\sigma_{\mathrm{EW}}$ are listed.}
\end{table}

Although the spectral resolution of the REOSC-DC data is insufficient for a detailed spectroscopic identification of the modes, the time series remains suitable for a time-based characterization of the variability in the line profile. In time-resolved spectroscopic studies, high-resolution data ($R \gtrsim 40\,000$) and high $S/Ns$ are typically preferred when the goal is to model subtle changes in the profile or derive oscillation quantum numbers \citep{Uytterhoeven2014}. However, for HD~75149, the moments $M_0$, $M_1$, and $M_3$ provide a robust diagnosis of whether the observed variability is dominated by changes in integrated line intensity, centroid velocity, or profile asymmetry. In this framework, variability that primarily affects $M_1$ is consistent with a nearly rigid shift of the profile, whereas simultaneous changes in $M_1$ and $M_3$ indicate genuine distortions of the line shape, as would be expected from photospheric velocity fields perturbed by non-radial pulsations or low-frequency stochastic variability \citep {Aerts2009,Asteroseismology2010}. 

\begin{figure}
    \centering
    \includegraphics[width=\linewidth]{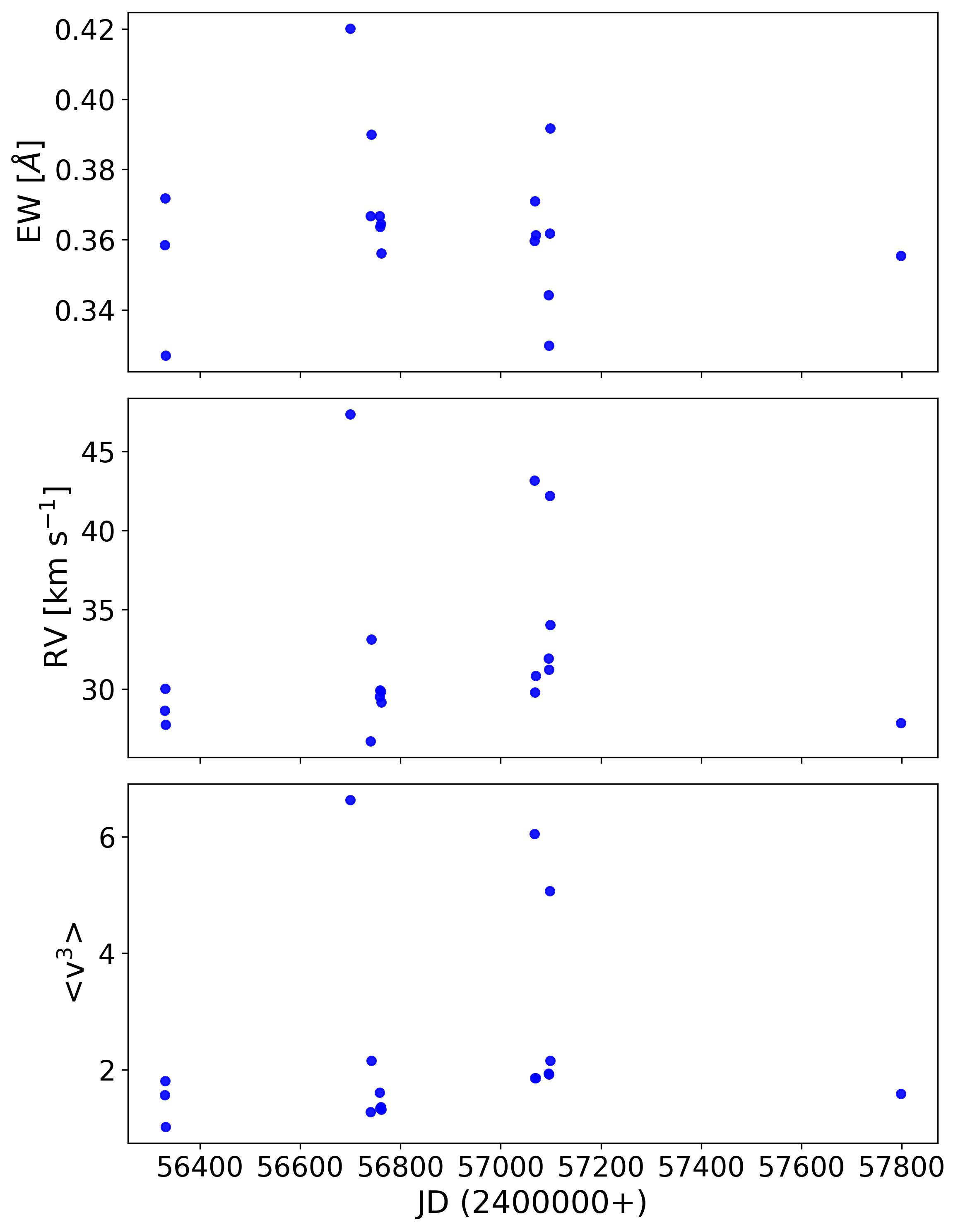}
    \caption{\textit{From top to bottom}:  EW ($M_0$), RV ($M_1$), and skewness $\langle v^{3} \rangle$ ($M_3$) for \ion{C}{ii} 6578.05 line. The measurements are concentrated in two main observing groups around 2014 and 2015. The EW values remain within a relatively narrow range compared to the RV and skewness, whereas the RV and $\langle v^{3} \rangle$ panels exhibit greater scatter and several high-value points in the central and later observing groups.}
    \label{fig:CIImoments}
\end{figure}

\begin{figure}
    \centering
    \includegraphics[width=\linewidth]{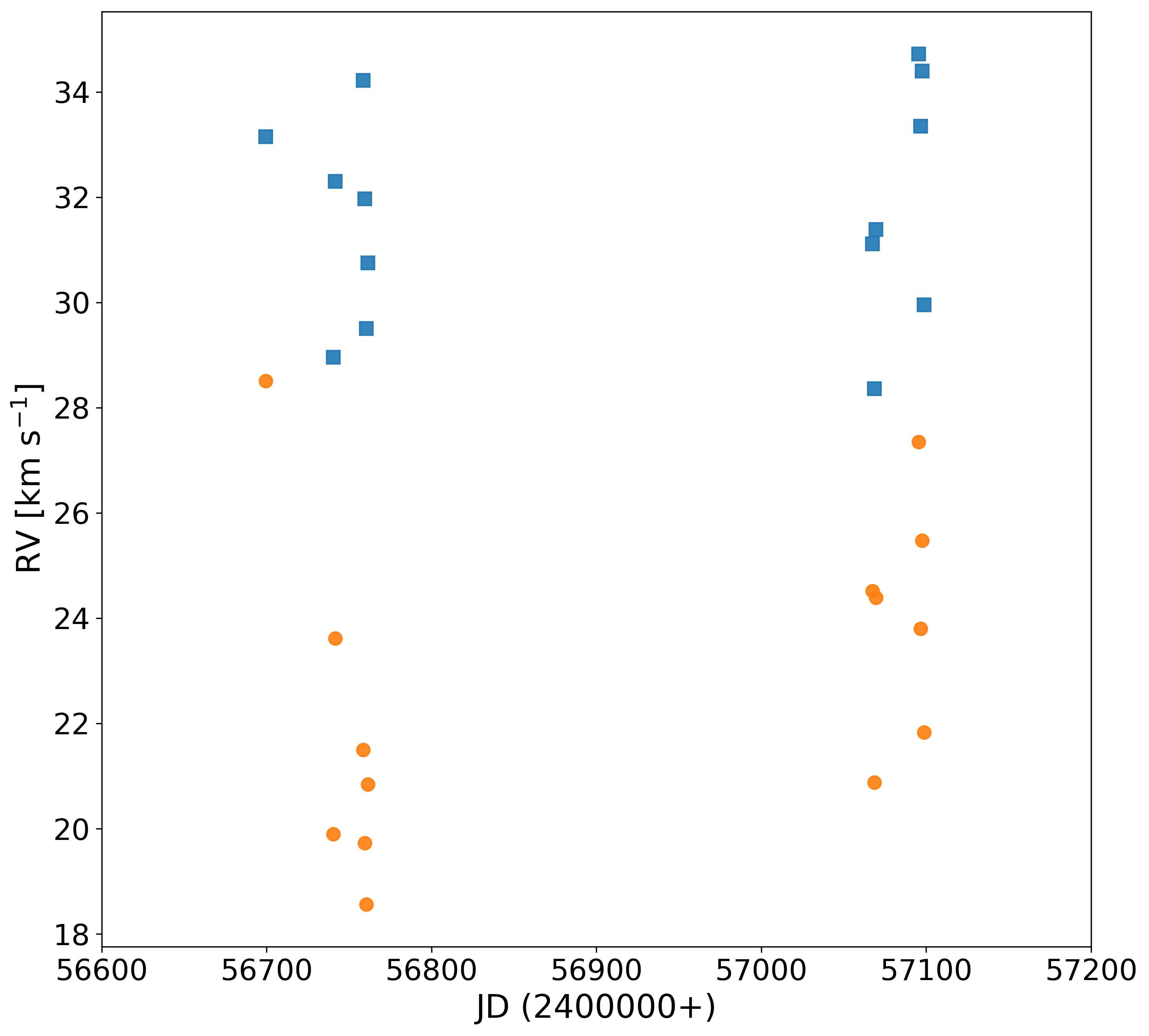}
    \caption{Absolute RVs for \ion{He}{i} 4471 (oranges circles) and \ion{Si}{iii} 4552 (blue squares) spectral lines. The data presented cover observations made in 2014 (March–April) and 2015 (February–March). Both lines show a similar temporal behavior, especially during the second period}.
    \label{fig:siiiicvheii}
\end{figure}

This is consistent with our results. As shown in Fig.~\ref{fig:CIImoments}, the \ion{C}{ii}\,$ $6578 line exhibits relatively limited variation in $M_0$, whereas $M_1$ and $M_3$ show the greatest variations in the central and rear observation groups, indicating that the variability cannot be explained as a simple rigid shift of a profile. This is also supported by the measured dispersions, since for \ion{C}{ii}\,$ $6578 we obtain $\sigma_{\rm RV}=5.57$ km s$^{-1}$ and $\sigma_{\rm EW}=0.04$~\AA, both of which are clearly above the average internal uncertainties ($\Delta {\rm RV}=0.49$ km s$^{-1}$ and $\Delta {\rm EW}=0.003$~\AA). In turn, Fig.~\ref{fig:siiiicvheii} shows that \ion{Si}{iii}\, 4552 and \ion{He}{i}\, 4471 follow a similar temporal pattern of RV, albeit with modest amplitudes, which supports a common photospheric origin rather than variability restricted to a single transition. Therefore, although REOSC-DC spectra do not allow for a formal mode-identification analysis, they do provide evidence of intrinsic photospheric variability, which is consistent with the line-profile variability commonly observed in BSGs and often interpreted in terms of pulsational or stochastic velocity fields, as in \citep{kraus2015}.


\subsection{Hydrogen lines}

Based on Table~\ref{tab:ha_var} for H$\alpha$, the absorption component shows RV variations between $10.26 \pm 1.04\,\,\rm km\,s^{-1}$ and $-146.81 \pm 1.60\,{\,\rm km\,s^{-1}}$, with very marked daily jumps in some trains on consecutive days (e.g., between JD~56760.50 and 56761.50 the change is $\sim\!29~\mathrm{km\,s^{-1}}$; between JD~57068.50 and 57069.50 it is $\sim\!22~\mathrm{km\,s^{-1}}$). The emission component has RVs between $\sim\!36$ and $133~\mathrm{km\,s^{-1}}$, with typical day-to-day variations of $\sim\!4$--$9~\mathrm{km\,s^{-1}}$. In EW, the absorption varies between $\sim\!0.01$ and $0.61$~\AA, while the emission is weak most of the time ($\mathrm{EW}_{\rm em}\approx -0.02$ to $-0.35$~\AA), except for a notable episode around JD~57097--57098 in which the emission intensifies to $\mathrm{EW}_{\rm em}\approx -0.89$ and -0.84~\AA,\ while the emission peak shifts to lower RVs (from $\sim\!43$ to $\sim\!36~\mathrm{km\,s^{-1}}$). In general, H$\alpha$ shows variability on scales of less than one day and up to a few days; due to the three to four day trains in different years, variability is also noted on scales of weeks.

From Table~\ref{tab:RV_var}, we obtain the daily RV variations for H$\beta$, which already reach moderate to large values in the earliest interval and remain relevant in the last: changes of $\sim\!24~\mathrm{km\,s^{-1}}$ in Feb 2013, $\sim\!6~\mathrm{km\,s^{-1}}$ in Mar 2014, and $\sim\!8.3~\mathrm{km\,s^{-1}}$ in 2025 (maximum in Feb 2013), while in the other 2014--2015 windows the typical variations are of only a few $\mathrm{km\,s^{-1}}$. From Table~\ref{tab:EW_var}, the EW of H$\beta$ changes daily by $\sim\!0.02$--$0.06$~\AA\ in time intervals with less activity, and it reaches its most significant excursions in Feb 2013 (peaks of $\sim\!0.54$~\AA) and Mar 2015 (peaks of $\sim\!0.14$~\AA), while the 2025 train shows comparatively smaller day-to-day changes ($\sim\!0.02$~\AA). In summary, H$\beta$ already displays substantial variability in the earliest interval, with episodic enhancements in 2015 rather than a monotonic increase toward the final time windows.
However, the variability in RV of H$\gamma$ is evident in Table~\ref{tab:RV_var}: daily changes of $\sim\!11.0~\mathrm{km\,s^{-1}}$ in 2013, $\sim\!4.6~\mathrm{km\,s^{-1}}$ in April 2014, and $\sim\!7.1~\mathrm{km\,s^{-1}}$ in March 2015 are recorded. In February 2015 and January 2025, the variations are smaller ($\sim\!0.3$--$3.3~\mathrm{km\,s^{-1}}$). For EW (Table~\ref{tab:EW_var}), H$\gamma$ is the most sensitive line: large daily increases from $\sim\!0.57$~\AA\ (Mar 2014) to $\sim\!0.60$~\AA\ (Apr 2014) are observed. Other notable variations are $\sim\!0.13$--$0.17$~\AA\ in 2015 and 2025.

The rapid response of H$\gamma$ is related to the physical mechanisms that control wind acceleration, clumping, or density structures, which can vary on short timescales due to pulsations or stochastic processes such as internal gravity waves (see Sect.~\ref{sec:5}). Since H$\gamma$ is sensitive to these conditions, it serves as an excellent diagnostic for transient phenomena such as episodes of wind strengthening or attenuation, revealing that the star's wind environment is highly dynamic.

\subsection{Helium lines}

Helium lines generally show small to moderate variations. \ion{He}{I}\, 4387 exhibits daily changes typically below $\sim\!3.5~\mathrm{km\,s^{-1}}$ (maximum close to $5.9~\mathrm{km\,s^{-1}}$). \ion{He}{I}\, 4471 is usually below $\sim\!4.5~\mathrm{km\,s^{-1}}$ but can reach higher values, with peaks of $\sim\!7.4~\mathrm{km\,s^{-1}}$ in the last period. \ion{He}{I}\, 5015 shows the largest RV excursions, occasionally reaching variations of up to $24.7~\mathrm{km\,s^{-1}}$, with other strong episodes of $\sim\!15$--$18~\mathrm{km\,s^{-1}}$, and otherwise remaining between $\sim\!0$ and $\sim\!6.1~\mathrm{km\,s^{-1}}$. \ion{He}{I}\, 5875 also shows significant RV variability, with values reaching up to $14.9~\mathrm{km\,s^{-1}}$ in the last period, in addition to episodes of $\sim\!3$--$6~\mathrm{km\,s^{-1}}$ in other time intervals.

In terms of EW, all helium lines show appreciable changes. \ion{He}{I}\, 4387 reaches 0.032~\AA. \ion{He}{I}\, 4471 reaches 0.089~\AA. \ion{He}{I}\, 5015 shows the greatest increase in EW, with a maximum of 0.200~\AA. \ion{He}{I}\, 5875 shows peaks of 0.114~\AA, and values that remain moderately high in the later epochs (up to 0.080~\AA\ in Mar 2015 and 0.073~\AA\ in Jan 2025). Overall, helium varies little in terms of RV, except in \ion{He}{I}\, 5015 and, to a lesser extent, \ion{He}{I}\, 5875.

\subsection{Silicon  lines}
The three silicon lines \ion{Si}{III}\,4552, 4567, and 4574 show low to moderate RV variations. The maxima are on the order of 1--4 km s$^{-1}$ for \ion{Si}{III}4552 (peak close to 4.4 km s$^{-1}$), up to 8.4 km s$^{-1}$ for \ion{Si}{III}\, 4567, and up to 5.1 km s$^{-1}$ for \ion{Si}{III}\, 4574. The core line variations are modest (typically below 0.02~\AA) with maxima of 0.022, 0.015, and 0.013~\AA\ for \ion{Si}{III}\,4552, 4567, and 4574, respectively (Table~\ref{tab:EW_var}). In summary, the \ion{Si}{III} lines do not show strong variations in either RV or EW, although \ion{Si}{III}\, 4567 displays the largest RV excursions among them (Table~\ref{tab:RV_var}).

To detect signs of binary activity, we specifically examined the profiles of the \ion{Si}{III} line, inspecting its radial velocities and line cores for any systematic shifts. These lines were selected because HD 75149 lies along a line of sight rich in H II regions (the Vela supernova remnant association), where nebular emission can contaminate stellar spectra; this especially affects the \ion{He}{I} lines \citep{ramirez2013}, making \ion{Si}{III} a cleaner diagnostic for binary-induced motion.

\subsection{Nitrogen lines}

\ion{N}{II}\, 4630 maintains mostly stable RV variation, with maxima up to $\sim\!6.4~\mathrm{km\,s^{-1}}$ (peak close to the same value). \ion{N}{II}\, 4601 shows moderate variations and reaches $6.2~\mathrm{km\,s^{-1}}$ in one of the periods. \ion{N}{II}\, 4607 shows slight variations; however, in 2015 there was a period of greater variation with a peak of $\sim\!10.5~\mathrm{km\,s^{-1}}$, although most of the time it remains at $\sim\!1$--$5~\mathrm{km\,s^{-1}}$. \ion{N}{II}\, 5666 is also active, with maxima reaching $\sim\!26.2~\mathrm{km\,s^{-1}}$.

The EW variations are slight but measurable: up to 0.017~\AA\ in \ion{N}{II}\,4601, up to 0.034~\AA\ in \ion{N}{II}\,4607, up to 0.014~\AA\ in \ion{N}{II}\,4630, and up to 0.040~\AA\ in \ion{N}{II}\, 5666. In general, \ion{N}{II}\,5666 is the most variable nitrogen line, while \ion{N}{II}\,4607 also shows clear but more moderate variability, and \ion{N}{II}\,4630 is among the most stable.

\subsection{Carbon lines}
The line of \ion{C}{ii}\, 6578 shows moderate to strong RV variations, ranging from sub-km\,s$^{-1}$ changes up to values above $10~\mathrm{km\,s^{-1}}$, with a maximum close to $13.4~\mathrm{km\,s^{-1}}$ (possibly because of its proximity to H$\alpha$). In contrast, \ion{C}{ii}\, 6582 shows clearly smaller RV variations, with a maximum close to $3.2~\mathrm{km\,s^{-1}}$ (and otherwise at the $\sim\!0.1$--$1.3~\mathrm{km\,s^{-1}}$ level depending on the period). The variation in EW reaches 0.045~\AA\ for \ion{C}{ii}\,6578 and up to 0.027~\AA\ for \ion{C}{II}\, 6582 (with typical values below 0.03~\AA).

\subsection{Magnesium lines}
The line \ion{Mg}{II}\, 4481 shows low to moderate changes in RV (maximum close to $6.3~\mathrm{km\,s^{-1}}$). However, its EW shows an increase in the April 2014 interval (see Table~\ref{tab:tracts}), with slight variations of $\sim\!0.08$~\AA, and lower levels the rest of the time (generally $\leq 0.05$~\AA).

\section{Stellar wind variability}\label{sec:4}
Several spectral lines, such as H$\alpha$, display a P-Cygni profile characterized by a blueshifted absorption component accompanied by a redshifted emission wing. This morphology arises from resonance scattering within an expanding envelope: material moving toward the observer absorbs continuum photons (producing blueshifted absorption), whereas material moving away re-emits photons (producing red emission). The presence and shape of P-Cygni profiles therefore provide direct diagnostics of the wind velocity field and the mass-loss rate \citep[e.g.,][]{cak1975,lamers1999}.

In this part, we derived the wind parameters (mass-loss rates and terminal velocities) by fitting the observed H$\alpha$ line profiles with synthetic lines. For this purpose, we made use of the ISOSCELES grid \citep{araya2024,araya2025}, a grid-based, quantitative, spectroscopic analysis of massive stars---which uses the Hydwind code \citep{cure2004}---to numerically resolve the hydrodynamic equations (momentum and mass conservation) that define an outflow fluid for non-fast-rotating BSGs; we also used the FASTWIND code \citep{Puls2005}, which is adapted to read the Hydwind output, thus enabling the 1D resolution of the radiative-transfer equation considering non-LTE effects to obtain the synthetic line profiles. Although the FASTWIND code can include microclumping, the ISOSCELES grid adopts $f_{\rm cl}=1$ (smooth wind; \citealt{araya2025}). Hence, the H$\alpha$-based values $\dot{M}$ reported here correspond to the smooth wind.
The wind-speed profile was constructed hydrodynamically and uses three line-force parameters: $\alpha$, $k$, and $\delta$. %, which govern the wind acceleration and determine the mass-loss rate and terminal velocity. 
The parameter $\alpha$ is a measure of the ratio between the line acceleration from optically thick lines and the total line acceleration (i.e., it sets the slope of the line-strength distribution), $k$ sets the overall line-strength normalization, and $\delta$ accounts for ionization effects throughout the wind. \citet{Cure2011} identified a new hydrodynamic regime, the $\delta$-slow solution, occurring for $\delta\gtrsim0.3$, which yields slower terminal velocities than the standard fast solution. Therefore, both fast and $\delta$-slow wind solutions are considered in the ISOSCELES grid. The theoretical spectra are characterized by stellar and wind parameters such as effective temperature ($\rm T_{eff}$), surface gravity ($\rm log\,g$), line-force parameters from the m-CAK theory, mass-loss rates, and terminal velocity; the latter is the final constant speed attained by material in the stellar wind. The grid also identifies the best-fitting synthetic profiles through a $\chi^{2}$ minimization scheme applied to one or several spectral lines, all with equal weight.

The ranges of line-force parameters are from 0.45 to 0.65 for $\alpha$, from 0.05 to 0.60 for $k$ and from 0.00 to 0.35 for $\delta$. Details of the steps are provided by \cite{araya2025} in their Table~1. For the stellar atmosphere models, the surface gravities range from $\log g\,$=\,4.3 down to approximately 90\% of the Eddington limit, with increments of 0.15 in the logarithms. The effective temperature ranges from $9\,000$ K to $45\,000$ K, using steps of 500 K below $30\,000$ K and $1\,000$ K above it.

Because the $\delta$-slow solution appears to better fit the observed features in BSGs with higher mass-loss rates and denser winds (\cite{Venero2024,araya2025}), we fixed the search method on the ISOSCELES grid to consider only this kind of solution. Because our goal is to quantify the mass-loss rate variation in consecutive days, the stellar parameters were set to $\log g$ = 2.5 dex and T$\mathrm{eff}$ = 16 \,500 K, following the values reported in the literature \citep{Haucke2018}, and we only modeled the H$\alpha$ line profile, which is the best optical line for measuring wind parameters. The best model was selected considering the lowest $\chi^2$ value. Although formal uncertainties are not explicitly derived here, the variations in the inferred mass-loss rates are significantly larger than the typical spacing of the ISOSCELES grid, indicating that the observed changes are robust against the resolution of the parameter space (see the discussion section).
For further details, see Appendix B of \cite{araya2025}.

\begin{figure*}
    \centering
    \includegraphics[width=0.45\linewidth]{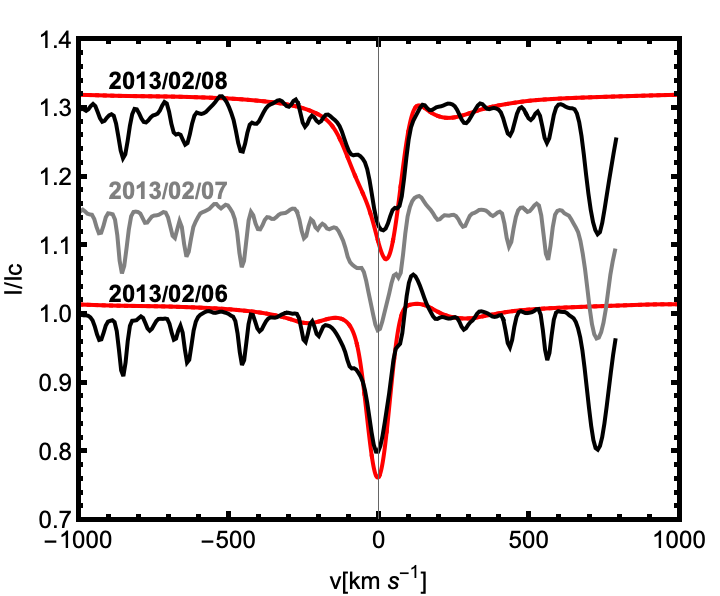}
    \includegraphics[width=0.45\linewidth]{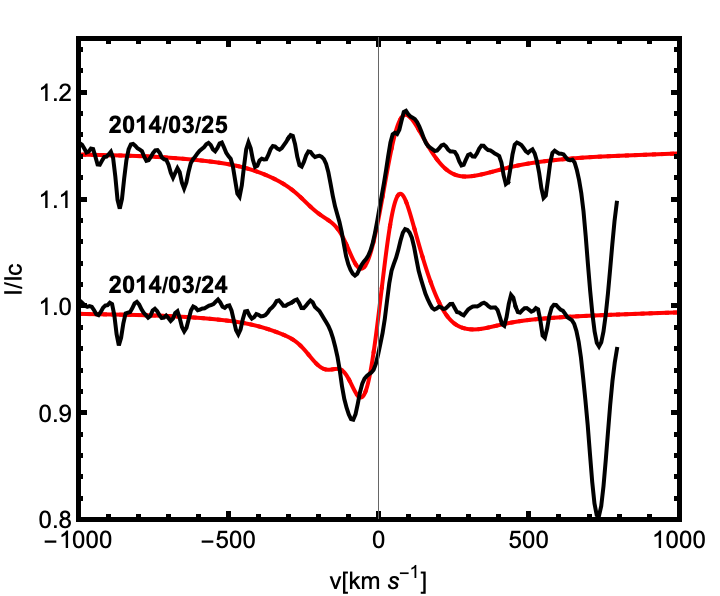}
    \includegraphics[width=0.45\linewidth]{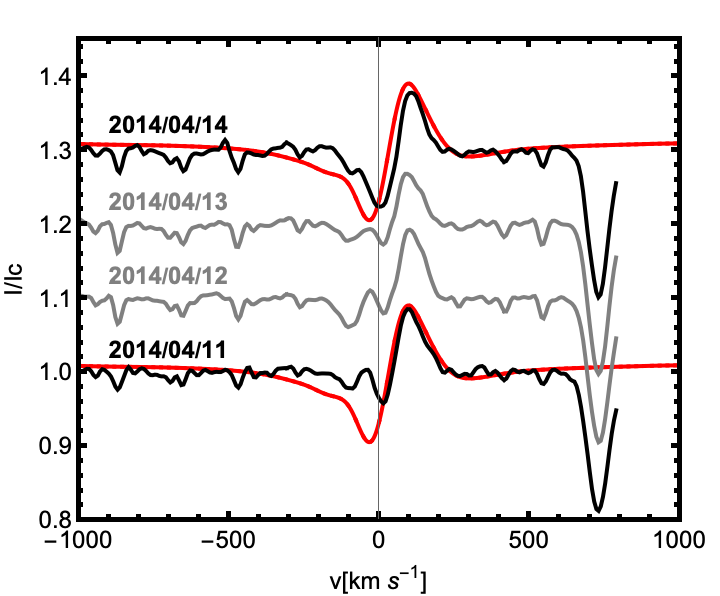}
    \includegraphics[width=0.45\linewidth]{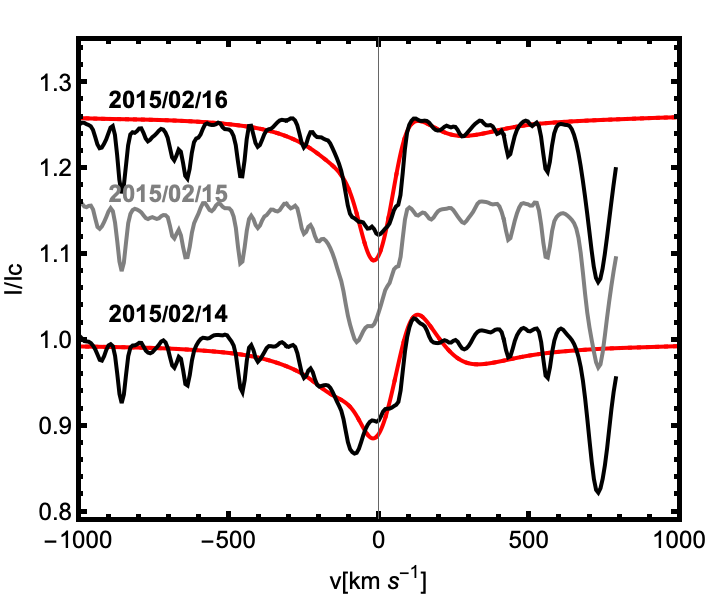}
    \includegraphics[width=0.45\linewidth]{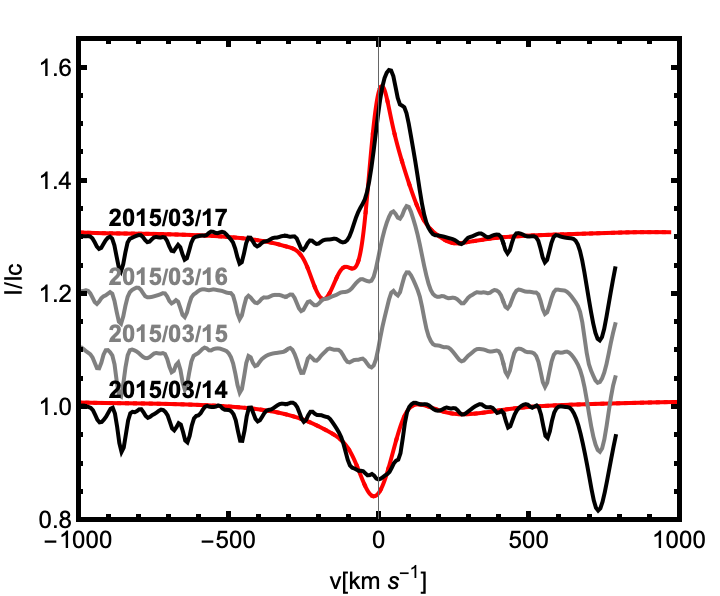}
    \includegraphics[width=0.45\linewidth]{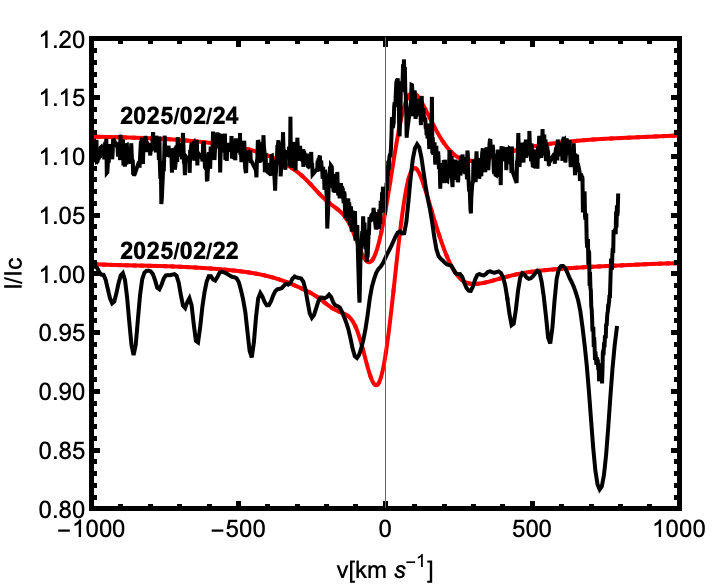}
    \caption{H$\alpha$ line profile variation of each time interval between 2013 and 2025. The solid black lines represent the fitted spectra, with the corresponding fits overplotted in red. Solid gray lines represent spectra within the time interval that were not used in the modeling.}
    \label{fig:model_fit}
\end{figure*}

To characterize the short-term variations on the order of days, we fitted the first and last spectra for each time interval. Figure~\ref{fig:model_fit} displays the fitted spectra in black and their corresponding best fit in red, and Table~\ref{tab:wind_parameters_unique} shows the corresponding wind parameters for each fit. 

\begin{table}[!ht]
    \centering
    \caption{Wind parameters obtained from spectra at time intervals with consecutive or close dates.}
    \begin{tabular}{cccccc}
    \hline
    \hline
        Obs date\rule{0pt}{2.6ex} & $\dot{\rm M}$  & $v_{\infty}$  & $\alpha$ & $k$ & $\delta$  \\ 
             & [$\rm 10^{-7}\, M_{\odot}\,yr^{-1}$] & [$\rm km\,s^{-1}$] &  &  &  \\    
        \hline
        2013-02-06 & 1.76 & 225.7 & 0.51 & 0.20 & 0.34 \\ 
        2013-02-08 & 1.06 & 251.9 & 0.53 & 0.15 & 0.35 \\ 
        \hline
        2014-03-24 & 2.39 & 263.8 & 0.55 & 0.15 & 0.35  \\ 
        2014-03-25 & 1.47 & 271.4 & 0.51 & 0.25 & 0.33  \\
        \hline
        2014-04-11 & 2.44 & 254.3 & 0.53 & 0.15 & 0.34 \\ 
        2014-04-14 & 2.44 & 254.3 & 0.53 & 0.15 & 0.34 \\ 
        \hline
        2015-02-14 & 1.47 & 271.4 & 0.51 & 0.25 & 0.33 \\ 
        2015-02-16 & 1.52 & 252.5 & 0.51 & 0.20 & 0.32 \\ 
        \hline
        2015-03-14 & 1.52 & 252.5 & 0.51 & 0.20 & 0.32 \\ 
        2015-03-17 & 2.81 & 248.6 & 0.51 & 0.20 & 0.33 \\ 
        \hline
        2025-01-22 & 2.44 & 254.3 & 0.53 & 0.15 & 0.34 \\ 
        2025-01-24 & 2.39 & 263.8 & 0.55 & 0.15 & 0.35 \\ 
        % \hline
        \hline
    \end{tabular}
    \label{tab:wind_parameters_unique}
\end{table}

The most drastic changes observed in the P-Cygni profiles are those for 2015. In February, the emission component decreased almost entirely over three days, and in March the line profile transitioned from absorption to emission over four days. In terms of the calculated mass-loss rate ($\rm\dot{M}$), significant changes occurred in March 2015, with the value increasing by a factor of 1.8. This increase in mass loss suggests that the star entered a phase of intense activity within two days, possibly related to phenomena such as stellar pulsations or instabilities in the stellar wind. Pulsation–wind coupling has also been reported for the BSG 55 Cyg, where multi-periodic nonradial pulsations correlate with cyclical changes in the H$\alpha$ morphology and mass-loss rate. The remaining time intervals show a decrease in the mass-loss rate, with the largest drop in $\dot{\rm M}$ of approximately 60\% (1.6 factor) occurring between two consecutive nights in March 2014 \citep[e.g.,][]{kraus2015,Cidale2023}.

In terms of $v_{\infty}$, its behavior seems to be opposite to the mass-loss value, i.e., when the mass loss increases, the terminal velocity decreases. However, we note that if in a fast wind (described by a $\beta$ law with $\beta \, \sim 0.8-1.2$) terminal-velocity changes can take more than a week, changes in a slow wind could take on the order of months\footnote{If $dr/dt=v_{\infty}(1-R_{\ast}/r)^{\beta}$, $\Delta t =v_{\infty}R_{\ast}\int_{1}^{\infty}{(1-1/x)^{-\beta}} $, with $x=r/R_{\ast}$}. Although, in our results the highest variation in terminal velocity is observed in 2013, with a variation of $\sim \rm 26 \, km\, s^{-1}$ over three days. In the rest of the models and for each time interval, the variation does not exceed $\rm 10 \, km\, s^{-1}$.

\section{Discussion}\label{sec:5}
The multi-epoch spectroscopic analyses presented in this work reveal that HD \,75149 exhibits pronounced short-term variability in its H$\alpha$ line. The combination of results from the line-by-line EW and RV analysis, along with wind modeling, indicates that the observed changes are primarily intrinsic to the star, rather than the result of binary motion.

On the one hand, with respect to the spectroscopic variability, the lack of significant RV variations in the metallic lines (\ion{Si}{III}, \ion{N}{II}, \ion{Mg}{II}, \ion{C}{II}) and the absence of coherent periodicity across epochs argue against the presence of a short-period binary companion. Instead, the correlated EW variability observed in the H, He, and Mg lines, particularly on timescales of days, suggests that photospheric or subsurface processes, such as stellar winds and/or radial pulsations, are at play. Compared with the literature, these short-timescale changes are consistent with the g- and s-mode oscillations typically seen in the $\alpha$ Cyg variables, which modulate the local temperature and density structure of the stellar photosphere (see, e.g., \citealt{kraus2015}). Additionally, the multi-periodic structure observed in TESS data in the literature, with frequencies corresponding to periods ranging from 1.1 to 10.4 days, further supports this interpretation. The absence of a single dominant frequency and the presence of several comparable modes indicate complex pulsational behavior, possibly involving coupling between multiple cavities in the stellar interior. This behavior resembles that of other well-studied $\alpha$ Cyg stars such as Deneb (\citealt{Abt2023}). However, in HD~75149, the pulsations appear as a continuous multiperiodic signal rather than isolated episodes, suggesting sustained excitation of multiple nonradial modes; this corresponds to the global correlation between the $M_1$ and $M_3$ spectral lines of the star. We note that our observational basis is limited: without evidence of strictly constant amplitude on longer timescales, this impression of greater stability remains tentative.
However, with respect to wind variability and $\delta$-slow solution, the H$\alpha$ line displays clear morphological changes, alternating among absorption, emission, and P-Cygni profiles on timescales of days to weeks. These variations reflect changes in wind density and ionization structure. By fitting the H$\alpha$ line with hydrodynamic models using $\delta$-slow solutions, we find that the $\delta$-slow regime systematically provides better agreement with observations, reproducing the lower terminal velocities and denser wind signatures typical of BSGs (\citealt{Venero2024,ortiz2025}). This solution arises when the ionization parameter, $\delta,$ increases, changing the balance between radiative acceleration and gravity in the external part of the atmosphere, which gives lower terminal velocities and higher densities in the wind base and matches the observed line profiles of BSGs.

The derived mass-loss rates vary between $1.06 \times 10^{-7}$ and $2.81 \times 10^{-7} \rm \,M_{\odot}\,yr^{-1}$, with terminal velocities between 225 and 271 $\rm km\,s^{-1}$. Variations of a factor of 1.8 are observed on fourth-day scales, an variations of a factor of 1.6 are seen on a one-day scale, confirming that HD\,75149 experiences rapid, dynamical wind fluctuations. Such short-term variability likely arises from pulsation–wind coupling, where photospheric oscillations perturb the base of the wind and modulate its density structure. Indeed, variability driven by pulsations has also been reported in other BSGs, such as 55 Cyg (\citealt{Cidale2023}). We acknowledge that the mass-loss variability inferred from our models cannot be quantified, as the uncertainties associated with the best-fit parameters have not yet been determined. Without these error estimates, the amplitude of the mass-loss changes remains formally unconstrained. Nevertheless, the rapid evolution of the P-Cygni line profile is clearly evident from visual inspection, indicating genuine short-term wind variability. Moreover, other works in the literature rarely determine uncertainties in wind parameters, because the typical method of searching for the best model is usually based on visual inspection. The ISOSCELES project uses a $\chi^2$ test to determine the best fits. 
The ISOSCELES team is planning to develop a Bayesian Markov Chain Monte Carlo (MCMC) method in the future to derive the uncertainties of the stellar and wind parameters. \\

The coexistence of multi-periodic pulsations and variable, slow winds places HD\,75149 within the transitional domain between stable BSGs and objects exhibiting enhanced mass loss. The success of the $\delta$-slow solution in reproducing the observed H$\alpha$ profiles reinforces its applicability for BSGs with slowly accelerating outflows. The results are also compatible with theoretical predictions of stochastic low-frequency variability in massive stars (\citealt{Bowman2019}). Internal gravity waves (IGWs) generated by subsurface convection zones (\citealt{Cantiello2021}) can coexist with coherent pulsation modes, resulting in the complex and irregular modulation of both photometric and spectroscopic signals. These waves may contribute to episodic increases in mass loss, which is consistent with observed day-scale changes in wind parameters.\\

Our findings confirm that HD\,75149 exhibits short-term spectroscopic variability that is consistent with changes in the mass-loss rate, and its wind properties are best reproduced by models adopting a $\delta$-slow hydrodynamic wind solution (with $\delta \simeq 0.3-0.35$; e.g., \citealt{Cure2011,Venero2024,araya2025}). BSGs with similarly dense, slow winds are expected to experience recurrent episodes of enhanced mass loss that efficiently remove angular momentum and strongly influence whether they subsequently evolve toward the red-supergiant branch or pass through a luminous blue variable phase \citep{Keszthelyi2017,Smith2014}.

In terms of the limitations of the study, consideration is given to the restricted continuous temporal coverage of the spectra, which skews the exhaustive survey of possible secular variations in the spectroscopic data. For this reason, future coordinated campaigns that combine high-resolution spectroscopy, space-based photometry, optical/infrared interferometry, and wind modeling will be necessary to validate and quantify the pulsation–wind connection in BSGs. Long-baseline interferometric observations can independently constrain fundamental parameters, such as the angular diameter, and spatially resolve extended atmospheres and winds at high angular and spectral resolution, allowing one to trace short-term changes in envelope geometry and density associated with instabilities and enhanced mass loss \citep[e.g.,][]{Chesneau2010}. In particular, time-dependent Non-LTE simulations could determine whether the rapid changes observed in HD\,75149 originate from mode beating, IGWs, or localized wind instabilities.

\section{Conclusions}\label{sec:6}
In this work, we conducted a multi-epoch spectroscopic study of the B3 Ia supergiant HD\,75149 (OP Vel), from 2006 to 2025, with 25 average optical spectra.
The spectroscopic variability of HD\,75149 appears to originate from intrinsic stellar processes affecting both the photosphere and the wind. The absence of coherent, orbit-like RV shifts in metallic lines rules out a close binary companion as the dominant direct driver of the variability. We cannot, however, completely exclude the possibility that a (wide or low-inclination) companion could indirectly contribute by exciting or enhancing non-radial pulsations through internal tidal mechanisms.

The H$\alpha$ profile exhibits substantial short-term changes between absorption, emission, and P-Cygni morphologies. The model that most accurately reproduces the observed profiles was obtained with a hydrodynamic solution, using a value of the $\delta$ line-force parameter in the range of 0.32--0.35, which corresponds to a $\delta$-slow solution. Variations on the order of two to four days point to rapid dynamical fluctuations in the stellar wind.

The observed correlation of timescales between wind and pulsational variability supports a pulsation–wind coupling scenario, where changes in photospheric structure modulate the mass-loss rate and terminal velocity. Finally, HD\,75149 probably has an intermediate evolutionary regime between relatively stable BSGs and objects exhibiting enhanced mass loss, in which nonradial pulsations, internal gravity waves, and wind instabilities interact. Understanding these processes is essential for constraining the pre-supernova evolution of massive stars.
These results highlight the importance of high-cadence spectroscopic monitoring to understand the short-term dynamics of massive-star winds and provide further observational support for the $\delta$-slow hydrodynamic regime predicted by line-driven wind theory.

The long-term baseline covered by our dataset illustrates the importance of coordinated campaigns with different observational techniques and consecutive epochs to characterize the variability of BSGs. Moreover, we expect to perform a high-cadence (consecutive days) spectroscopic campaign on more BSGs since more of them have monthly or yearly observations, making it impossible to determine short-term pulsational cycles and slower secular changes in the wind by spectroscopic observations. It will be imperative to place HD\,75149 and similar objects into a broader, statistically meaningful framework.
\section*{Data availability}
Tables~\ref{tab:ha_var}, ~\ref{tab:RV_var}, and~\ref{tab:EW_var} and the reduced spectra are also available in electronic form at the CDS via anonymous ftp to cdsarc.u-strasbg.fr (130.79.128.5) or via \url{http://cdsweb.u-strasbg.fr/cgi-bin/qcat?J/A+A/}.

\begin{acknowledgements}
JC is giving her warm thanks to the Universidad de Antofagasta, the Graduate School of the same university, and the Regional Government of the Antofagasta Region, through the project “Transfer and strengthening of the doctoral program in Astrophysics and Astroinformatics, BIP 40067595” for the period 2025. JC \& CA thanks for the support from the FONDECYT project 11190945. MC, IA \& CA thank the support from ANID FONDECYT projects 1230131 and 1261498. MC and CA acknowledge partial support from Centro de Astrofísica de Valparaíso. NM thanks the support from ANID BECAS / DOCTORADO NACIONAL 21221364. This work has been made possible thanks to the use of AWS-U.Chile-NLHPC credits. Powered@NLHPC: This research was partially supported by the NLHPC's supercomputing infrastructure (ECM-02) and co-funded by the European Union (project 101183150 - OCEANS).
\end{acknowledgements}

\bibliographystyle{aa}
\bibliography{references}

\raggedbottom  
\clearpage
\onecolumn

\begin{appendix} 
\section{RV and EW measurements}\label{appendixA}
In this appendix we present the measurements of RV and equivalent width for all lines analyzed in the star.

\begin{table}[ht]
\centering
\caption{RVs and EWs observed in the H$\alpha$ line.}
\label{tab:ha_var}
\begin{tabular}{r c c c c}
\hline
\hline
JD & RV$_{\rm abs}$ [km s$^{-1}$] & EW$_{\rm abs}$ [\AA] & RV$_{\rm em}$ [km s$^{-1}$] & EW$_{\rm em}$ [\AA]\\
\hline
53367.87 & $-13.51 \pm 0.47$ & $0.185 \pm 0.004$ & $117.73 \pm 1.24$ & $-0.141 \pm 0.004$ \\
53484.59 & $-0.82 \pm 1.45$ & $0.325 \pm 0.005$ & $112.98 \pm 1.00$ & $-0.057 \pm 0.004$ \\
53484.59 & $0.59 \pm 1.04$ & $0.322 \pm 0.004$ & $112.82 \pm 0.98$ & $-0.056 \pm 0.004$ \\
53740.75 & $-37.90 \pm 0.70$ & $0.256 \pm 0.002$ & $132.60 \pm 3.03$ & $-0.093 \pm 0.001$ \\
53773.79 & $-5.01 \pm 0.61$ & $0.506 \pm 0.003$ &  &  \\
53773.79 & $-5.26 \pm 0.59$ & $0.510 \pm 0.003$ &  &  \\
56329.50 & $-45.32 \pm 0.15$ & $0.549 \pm 0.000$ & $119.76 \pm 0.60$ & $-0.066 \pm 0.001$ \\
56330.65 & $-23.06 \pm 0.85$ & $0.509 \pm 0.003$ & $128.93 \pm 1.81$ & $-0.023 \pm 0.001$ \\
56331.67 & $0.22 \pm 0.08$ & $0.557 \pm 0.006$ &  &  \\
56699.50 &  &  & $67.25 \pm 3.63$ & $-0.322 \pm 0.005$ \\
56740.58 & $-82.37 \pm 0.51$ & $0.261 \pm 0.001$ & $88.71 \pm 0.12$ & $-0.134 \pm 0.000$ \\
56741.59 & $-76.45 \pm 0.45$ & $0.355 \pm 0.001$ & $95.35 \pm 0.33$ & $-0.052 \pm 0.000$ \\
56758.50 & $5.95 \pm 0.59$ & $0.038 \pm 0.001$ & $112.67 \pm 0.26$ & $-0.173 \pm 0.000$ \\
56759.50 & $10.26 \pm 1.04$ & $0.010 \pm 0.000$ & $111.99 \pm 0.19$ & $-0.183 \pm 0.000$ \\
56760.50 & $4.49 \pm 0.75$ & $0.021 \pm 0.002$ & $107.61 \pm 0.19$ & $-0.132 \pm 0.001$ \\
56761.50 & $-24.95 \pm 0.48$ & $0.173 \pm 0.001$ & $116.16 \pm 0.30$ & $-0.139 \pm 0.001$ \\
57067.50 & $-77.45 \pm 0.31$ & $0.595 \pm 0.001$ & $126.40 \pm 0.44$ & $-0.028 \pm 0.000$ \\
57068.50 & $-62.35 \pm 0.04$ & $0.607 \pm 0.001$ &  &  \\
57069.50 & $-40.48 \pm 0.31$ & $0.547 \pm 0.001$ &  &  \\
57095.50 & $-62.80 \pm 1.20$ & $0.046 \pm 0.001$ & $86.84 \pm 0.33$ & $-0.349 \pm 0.001$ \\
57096.50 &  &  & $59.46 \pm 0.26$ & $-0.482 \pm 0.002$ \\
57097.50 &  &  & $42.61 \pm 0.60$ & $-0.891 \pm 0.002$ \\
57098.63 &  &  & $35.81 \pm 1.73$ & $-0.842 \pm 0.009$ \\
57798.50 & $-146.81 \pm 1.60$ & $0.176 \pm 0.003$ & $99.85 \pm 0.37$ & $-0.235 \pm 0.002$ \\
60698.36 & $-81.28 \pm 1.38$ & $0.186 \pm 0.003$ & $81.20 \pm 0.61$ & $-0.138 \pm 0.002$ \\
60698.36 & $-80.36 \pm 1.66$ & $0.185 \pm 0.003$ & $80.57 \pm 0.51$ & $-0.137 \pm 0.002$ \\
60700.72 & $-64.72 \pm 1.87$ & $0.320 \pm 0.002$ & $90.38 \pm 0.42$ & $-0.175 \pm 0.002$ \\
60700.73 & $-62.41 \pm 0.86$ & $0.325 \pm 0.002$ & $90.26 \pm 0.30$ & $-0.177 \pm 0.002$ \\
\hline
\end{tabular}
\tablefoot{In cases where a P-Cygni profile is present, we have RV and EW values for both emission and absorption. Unlike cases where a purely emission or absorption profile is observed, in these cases, the opposite option remains empty.}
\end{table}

\begin{figure}[h!]
        \centering
        \includegraphics[width=0.9\textwidth]{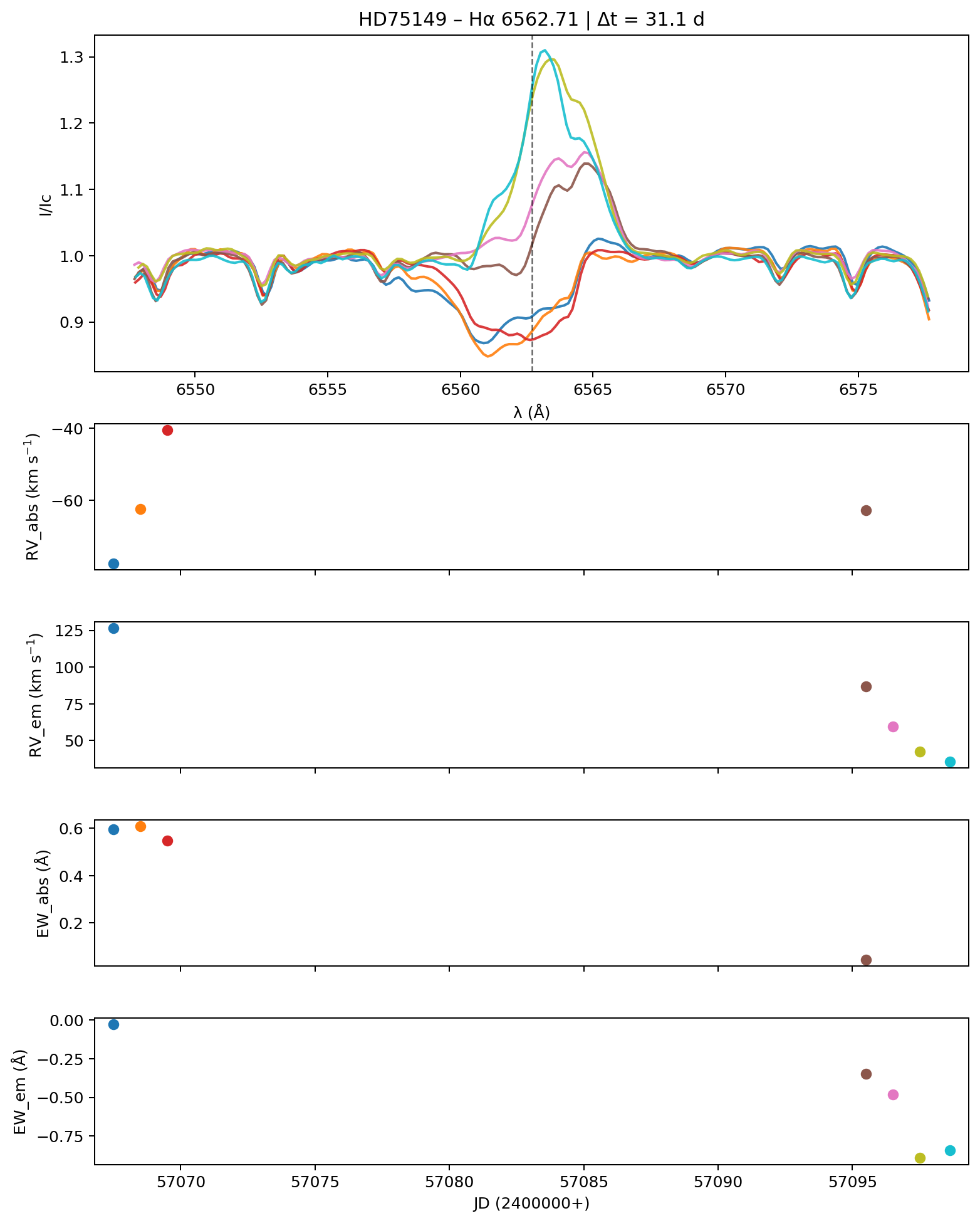}
        \caption{RV and EW of absorption and emission component of the P-Cygni H$\alpha$ profile. We consider an error of 3 km s$^{-1}$ for RV and 0.02-0.03 [$\AA$] for the EW.}
        \label{fig:rvha}
 \end{figure}    

\newcommand{\nodata}{--}         
\newcommand{\sig}[1]{#1} 

\begin{table}[htbp]
\centering
\caption{RV variation in spectral lines.}
\label{tab:RV_var}
\begin{tabular}{c|rr|c|rrr|rr|rrr|c}
\hline\hline
& \multicolumn{12}{c}{RV Variation [km s$^{-1}$]} \\
\hline
Line
 & \multicolumn{2}{c}{Feb 2013}
& \multicolumn{1}{c}{Mar 2014}
& \multicolumn{3}{c}{Apr 2014}
& \multicolumn{2}{c}{Feb 2015}
& \multicolumn{3}{c}{Mar 2015}
& \multicolumn{1}{c}{Jan 2025} \\
 & (1) & (2) & (1) & (1) & (2) & (3) & (1) & (2) & (1) & (2) & (3) & (1) \\
\cline{2-3}\cline{4-4}\cline{5-7}\cline{8-9}\cline{10-12}\cline{13-13}
\hline\hline
H$\gamma$            & \sig{10.973} & \sig{6.247} & \sig{3.496} & \sig{2.845} & \sig{0.036} & \sig{4.603} & \sig{1.378} & \sig{3.257} & \sig{0.019} & \sig{0.766} & \sig{7.115} & \sig{0.345} \\
\ion{He}{I}4387      & \sig{3.367} & \sig{0.760} & \sig{5.713} & \sig{3.194} & \sig{0.484} & \sig{0.565} & \sig{5.930} & \sig{1.515} & \sig{3.456} & \sig{3.241} & \sig{0.428} & \sig{0.879} \\
\ion{He}{I}4471      & \sig{2.392} & \sig{4.450} & \sig{3.717} & \sig{1.768} & \sig{1.167} & \sig{2.278} & \sig{3.634} & \sig{3.514} & \sig{3.549} & \sig{1.673} & \sig{3.641} & \sig{7.419} \\
\ion{Mg}{II}4481     & \sig{3.804} & \sig{0.156} & \sig{6.270} & \sig{0.452} & \sig{1.655} & \sig{2.012} & \sig{5.136} & \sig{5.338} & \sig{2.725} & \sig{3.381} & \sig{0.433} & \sig{4.470} \\
\ion{Si}{III}4552    & \sig{2.985} & \sig{4.330} & \sig{3.336} & \sig{2.251} & \sig{2.461} & \sig{1.247} & \sig{2.757} & \sig{3.027} & \sig{1.368} & \sig{1.048} & \sig{4.446} & \sig{0.464} \\
\ion{Si}{III}4567    & \sig{0.739} & \sig{0.381} & \sig{5.494} & \sig{0.165} & \sig{2.436} & \sig{3.190} & \sig{6.863} & \sig{1.452} & \sig{2.275} & \sig{2.917} & \sig{8.398} & \sig{2.401} \\
\ion{Si}{III}4574    & \sig{0.544} & \sig{0.254} & \sig{5.141} & \sig{5.050} & \sig{3.120} & \sig{2.636} & \sig{0.461} & \sig{2.220} & \sig{2.277} & \sig{1.107} & \sig{3.566} & \nodata \\
\ion{N}{II}4601      & \sig{3.995} & \sig{0.700} & \sig{3.155} & \sig{1.511} & \sig{1.223} & \sig{6.200} & \sig{2.517} & \sig{0.878} & \sig{0.836} & \sig{2.659} & \sig{1.240} & \sig{2.659} \\
\ion{N}{II}4607      & \sig{3.797} & \sig{0.922} & \sig{0.311} & \sig{3.419} & \sig{1.024} & \sig{5.432} & \sig{10.477} & \sig{5.964} & \sig{9.744} & \sig{5.069} & \sig{1.537} & \nodata \\
\ion{N}{II}4630      & \sig{1.764} & \sig{0.733} & \sig{6.448} & \sig{3.244} & \sig{1.362} & \sig{3.652} & \sig{4.922} & \sig{1.689} & \sig{3.111} & \sig{0.679} & \sig{1.515} & \sig{1.129} \\
H$\beta$             & \sig{24.379} & \sig{3.098} & \sig{5.689} & \sig{2.444} & \sig{1.831} & \sig{1.650} & \sig{4.161} & \sig{4.305} & \sig{1.543} & \sig{1.415} & \sig{0.260} & \sig{8.303} \\
\ion{He}{I}5015      & \sig{6.139} & \sig{1.169} & \sig{15.002} & \sig{0.007} & \sig{0.882} & \sig{24.691} & \sig{4.055} & \sig{4.873} & \sig{15.008} & \sig{18.089} & \sig{10.103} & \sig{10.000} \\
\ion{N}{II}5666      & \sig{1.909} & \sig{3.020} & \sig{26.233} & \sig{0.872} & \sig{5.057} & \sig{8.712} & \sig{0.361} & \sig{3.078} & \sig{5.796} & \sig{3.856} & \sig{12.399} & \sig{1.887} \\
\ion{He}{I}5875      & \sig{2.624} & \sig{2.640} & \sig{3.086} & \sig{6.136} & \sig{2.898} & \sig{0.018} & \sig{5.195} & \sig{3.998} & \sig{0.386} & \sig{0.415} & \sig{2.588} & \sig{14.902} \\
\ion{C}{II}6578      & \sig{1.379} & \sig{2.271} & \sig{6.433} & \sig{0.410} & \sig{0.093} & \sig{0.669} & \sig{13.369} & \sig{1.045} & \sig{0.702} & \sig{10.977} & \sig{8.153} & \sig{4.253} \\
\ion{C}{II}6582      & \sig{1.188} & \sig{2.885} & \sig{3.172} & \sig{1.034} & \sig{0.908} & \sig{0.815} & \sig{1.233} & \sig{3.179} & \sig{0.081} & \sig{1.169} & \sig{1.303} & \nodata \\

\hline
\end{tabular}
\tablefoot{Each entry gives the day-to-day RV variation within the corresponding time window (see Table~\ref{tab:tracts}). Subcolumns (1), (2), (3) follow the internal subdivision used in that window. Missing measurements are indicated by non-data.}
\end{table}

\begin{table}[htbp]
\centering
\caption{EW variation in spectral lines.}
\label{tab:EW_var}
\begin{tabular}{c|rr|c|rrr|rr|rrr|c}
\hline\hline
& \multicolumn{12}{c}{EW Variation [\AA]} \\
\hline
Line
 & \multicolumn{2}{c}{Feb 2013}
& \multicolumn{1}{c}{Mar 2014}
& \multicolumn{3}{c}{Apr 2014}
& \multicolumn{2}{c}{Feb 2015}
& \multicolumn{3}{c}{Mar 2015}
& \multicolumn{1}{c}{Jan 2025} \\
 & (1) & (2) & (1) & (1) & (2) & (3) & (1) & (2) & (1) & (2) & (3) & (1) \\
\cline{2-3}\cline{4-4}\cline{5-7}\cline{8-9}\cline{10-12}\cline{13-13}
\hline\hline
H$\gamma$            & \sig{0.113} & \sig{0.144} & \sig{0.565} & \sig{0.024} & \sig{0.111} & \sig{0.604} & \sig{0.029} & \sig{0.007} & \sig{0.164} & \sig{0.131} & \sig{0.163} & \sig{0.166} \\
\ion{He}{I}4387      & \sig{0.025} & \sig{0.010} & \sig{0.011} & \sig{0.010} & \sig{0.011} & \sig{0.006} & \sig{0.002} & \sig{0.009} & \sig{0.002} & \sig{0.028} & \sig{0.032} & \sig{0.006} \\
\ion{He}{I}4471      & \sig{0.012} & \sig{0.020} & \sig{0.077} & \sig{0.017} & \sig{0.022} & \sig{0.019} & \sig{0.025} & \sig{0.020} & \sig{0.015} & \sig{0.040} & \sig{0.089} & \sig{0.022} \\
\ion{Mg}{II}4481     & \sig{0.010} & \sig{0.017} & \sig{0.054} & \sig{0.016} & \sig{0.011} & \sig{0.077} & \sig{0.013} & \sig{0.013} & \sig{0.008} & \sig{0.018} & \sig{0.043} & \sig{0.015} \\
\ion{Si}{III}4552    & \sig{0.001} & \sig{0.004} & \sig{0.022} & \sig{0.011} & \sig{0.004} & \sig{0.010} & \sig{0.002} & \sig{0.014} & \sig{0.001} & \sig{0.003} & \sig{0.022} & \sig{0.012} \\
\ion{Si}{III}4567    & \sig{0.004} & \sig{0.014} & \sig{0.005} & \sig{0.005} & \sig{0.001} & \sig{0.001} & \sig{0.012} & \sig{0.006} & \sig{0.015} & \sig{0.009} & \sig{0.003} & \sig{0.003} \\
\ion{Si}{III}4574    & \sig{0.007} & \sig{0.005} & \sig{0.007} & \sig{0.013} & \sig{0.005} & \sig{0.001} & \sig{0.003} & \sig{0.007} & \sig{0.007} & \sig{0.011} & \sig{0.009} & \nodata \\
\ion{N}{II}4601      & \sig{0.017} & \sig{0.011} & \sig{0.017} & \sig{0.009} & \sig{0.004} & \sig{0.013} & \sig{0.008} & \sig{0.004} & \sig{0.013} & \sig{0.001} & \sig{0.004} & \sig{0.001} \\
\ion{N}{II}4607      & \sig{0.020} & \sig{0.014} & \sig{0.008} & \sig{0.002} & \sig{0.004} & \sig{0.001} & \sig{0.016} & \sig{0.001} & \sig{0.034} & \sig{0.002} & \sig{0.002} & \nodata \\
\ion{N}{II}4630      & \sig{0.008} & \sig{0.003} & \sig{0.004} & \sig{0.013} & \sig{0.010} & \sig{0.014} & \sig{0.010} & \sig{0.004} & \sig{0.014} & \sig{0.004} & \sig{0.001} & \sig{0.001} \\
H$\beta$             & \sig{0.537} & \sig{0.083} & \sig{0.085} & \sig{0.030} & \sig{0.054} & \sig{0.044} & \sig{0.113} & \sig{0.005} & \sig{0.124} & \sig{0.056} & \sig{0.135} & \sig{0.020} \\
\ion{He}{I}5015      & \sig{0.085} & \sig{0.036} & \sig{0.072} & \sig{0.011} & \sig{0.047} & \sig{0.200} & \sig{0.028} & \sig{0.011} & \sig{0.051} & \sig{0.101} & \sig{0.075} & \sig{0.045} \\
\ion{N}{II}5666      & \sig{0.018} & \sig{0.016} & \sig{0.028} & \sig{0.023} & \sig{0.007} & \sig{0.026} & \sig{0.006} & \sig{0.016} & \sig{0.031} & \sig{0.030} & \sig{0.040} & \sig{0.006} \\
\ion{He}{I}5875      & \sig{0.022} & \sig{0.003} & \sig{0.114} & \sig{0.052} & \sig{0.043} & \sig{0.069} & \sig{0.041} & \sig{0.011} & \sig{0.039} & \sig{0.072} & \sig{0.080} & \sig{0.073} \\
\ion{C}{II}6578      & \sig{0.013} & \sig{0.045} & \sig{0.023} & \sig{0.003} & \sig{0.001} & \sig{0.008} & \sig{0.011} & \sig{0.010} & \sig{0.014} & \sig{0.032} & \sig{0.030} & \sig{0.001} \\
\ion{C}{II}6582      & \sig{0.006} & \sig{0.027} & \sig{0.005} & \sig{0.013} & \sig{0.005} & \sig{0.008} & \sig{0.003} & \sig{0.005} & \sig{0.013} & \sig{0.017} & \sig{0.017} & \nodata \\

\hline
\end{tabular}
\tablefoot{Each entry gives the day-to-day EW variation within the corresponding time window (see Table~\ref{tab:tracts}). Subcolumns (1), (2), (3) follow the internal subdivision used in that window. Missing measurements are indicated by non-data.}
\end{table}

\end{appendix}

\end{document}